\documentclass{osa-article}

\journal{oe}

\begin{document}

\noindent© 2022 Optica Publishing Group. Users may use, reuse, and build upon the article, or use the article for text or data mining, so long as such uses are for non-commercial purposes and appropriate attribution is maintained. All other rights are reserved. This work has been published in Optics Express with DOI:/10.1364/OE.444063.

\newpage

\title{Microring resonators with external optical feedback for time delay reservoir computing}

\author{Giovanni Donati\authormark{1,2}, Claudio R. Mirasso \authormark{1}, Mattia Mancinelli\authormark{2}, Lorenzo Pavesi\authormark{2} and Apostolos Argyris\authormark{1}}

\address{
\authormark{1}Instituto de Física Interdisciplinar y Sistemas Complejos, IFISC (CSIC-UIB), Campus UIB, 07122 Palma de Mallorca, Spain\\
\authormark{2}Nanoscience Laboratory, Department of Physics, University of Trento, Via Sommarive 14, 38123 Trento, Italy\\
}

\email{giovanni@ifisc.uib-csic.es} 


\begin{abstract}
Microring resonators (MRRs) are a key photonic component in integrated devices, due to their small size, low insertion losses, and passive operation. While the MRRs have been established for optical filtering in wavelength-multiplexed systems, the nonlinear properties that they can exhibit give rise to new perspectives on their use. For instance, they have been recently considered for introducing optical nonlinearity in photonic reservoir computing systems. In this work, we present a detailed numerical investigation of a silicon MRR operation, in the presence of external optical feedback, in a time delay reservoir computing scheme. We demonstrate the versatility of this compact, passive device, by exploiting different operating regimes and solving computing tasks with diverse memory requirements. We show that when large memory is required, as it occurs in the Narma 10 task, the MRR nonlinearity does not play a significant role when the photodetection nonlinearity is involved, while the contribution of the external feedback is significant. On the contrary, for computing tasks such as the Mackey-Glass and the Santa Fe chaotic timeseries prediction, the MRR and the photodetection nonlinearities contribute both to efficient computation. The presence of optical feedback improves the prediction of the Mackey-Glass timeseries while plays a minor role in the Santa Fe timeseries case.
\end{abstract}


\section{Introduction}
Microring resonators (MRRs) have emerged as fundamental building blocks in photonics due to their compact footprint, high bandwidth, large third-order nonlinearities, and the possibility to be integrated into various material substrates. They are used in a range of applications \cite{Heebner}, exploiting their resonance nature, including optical filtering \cite{Little1997}, optical switching \cite{Cheng2018}, optical sensing \cite{mesaritakis2010,Sensing2019} and complex integration of optical signals \cite{Ferrara2010}. The common MRR operation in these applications is to tune the resonance wavelength, by modifying its refractive index. This change can be triggered in different ways. For example, by electro-optic modulation \cite{Xu2005}, where a p-i-n junction is embedded in the ring and operates in an alternating forward and reverse bias condition, or by engineering the ring surface with an adsorbed layer for specific external detection \cite{Sensing2019}. Another possibility is to vary the temperature of the MRR waveguide, exploiting its thermo-optic coefficient \cite{Zhang2013}. Variations of the refractive index in silicon MRRs can be also induced by the propagating optical signal when the optical power circulating the device is high enough to activate two-photon absorption (TPA) and free-carrier dispersion (FCD) \cite{Johnson2006}. In this case, additional free carriers and phonons are generated, resulting in a passive nonlinear operation. These nonlinear effects have been explored in applications such as memory storage \cite{Almeida2004}, all-optical modulation \cite{Preble2005} and all-optical logic operations \cite{Xu2007}. Recently, MRRs were considered in neuromorphic photonics as promising candidates for optical nodes in computing structures, due to several dynamical features common to biological neurons such as self pulsations, excitability, and inhibitory spiking behavior \cite{Vaerenbergh2012,Xiang2020}. They were also proposed as nonlinear optical elements in photonic reservoir computing (RC) concepts \cite{ESN_2004,LSM_2002}, in integrated platforms \cite{MR_RC} and for applications in transmission channel equalization \cite{Li2021}.

In this work, we investigate and evaluate the performance of MRRs in the context of a time delay RC, a simplification of the RC concept which was initially introduced in \cite{Appeltant_2011}. In this approach, only one nonlinear (real) node is used to emulate the reservoir, in presence of a time delay that introduces recurrent connectivity between time-delayed node responses. In this way, a set of virtual nodes can be defined in a time-multiplexed form. Several implementations adopting this approach have been numerically and experimentally investigated, including electronic and optoelectronic devices \cite{Appeltant_2011,Larger2012}, all-optical devices \cite{Brunner_Laser,laserNumeric2014,Bueno2017} and also photonic integrated circuits based on semiconductor lasers with optical feedback \cite{Takano2018,Harkhoe2020}. Recently, a single silicon MRR in absence of feedback has also shown the potentiality to solve memory demanding tasks, based on its own nonlinear memory and virtual nodes defined by time multiplexing \cite{Borghi_MRR}. In this study, we specifically consider a silicon MRR subject to delayed optical feedback and we exploit the free carrier nonlinearity effects while introducing memory through the external cavity. We investigate the computational consistency and the memory properties of the overall system, by directly evaluating its performance on benchmark tasks that are known to require different compromises between memory and nonlinearity. Our findings, which are precursory to an experimental investigation, suggest that both the MRR and the external feedback are memory sources of the system. For computing tasks that demand memory that exceeds the one provided by the MRR, we show that the external feedback significantly improves the computing performance. The manuscript is structured as follows: the model of the MRR in presence of external optical feedback is described in section \ref{sec_Model}. In section \ref{Sec_RCscheme} we describe its implementation in a time-delay RC scheme and finally, in section \ref{Sec_results} we present and analyze the results obtained for the Narma 10, Mackey-Glass, and Santa Fe chaotic timeseries prediction tasks.

\section{MRR with optical feedback}
\begin{figure}[htbp]
\centering\includegraphics[scale=0.3]{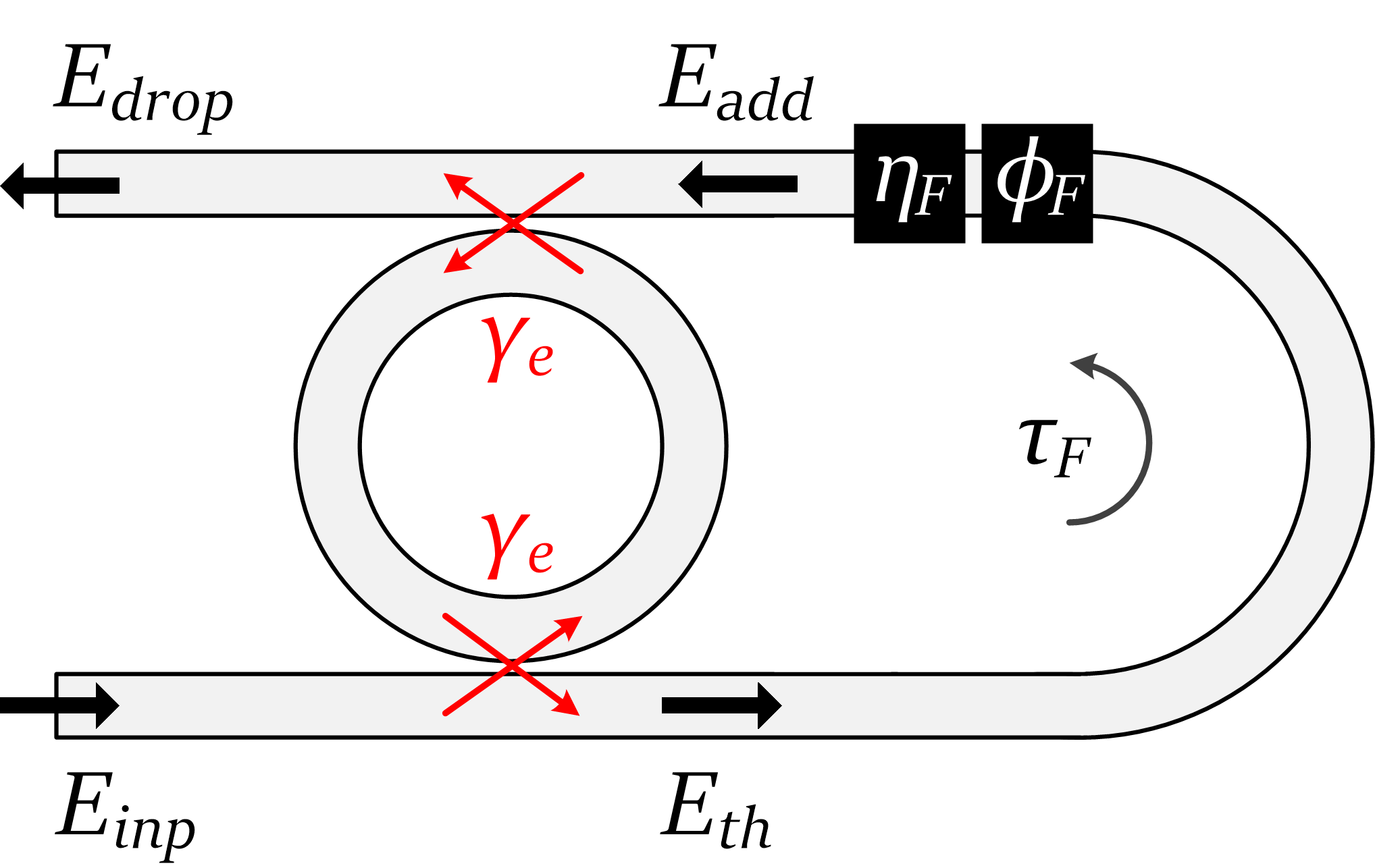}
\caption{MRR structure in an add-drop filter configuration with external optical feedback. $\gamma_e$ represents the MRR extrinsic losses due to the coupling with the straight waveguides, while $E_{inp}$, $E_{th}$, $E_{add}$ and $E_{drop}$ represent the electric field amplitudes in the correspondent ports.}
\label{Ring_theory}
\end{figure}

The MRR structure in presence of external optical feedback is illustrated in Fig. (\ref{Ring_theory}). It operates in an add-drop filter configuration, with symmetric coupling in the interconnection with the two straight waveguides. It receives the input signal $E_{inp}$ from the input port and provides an output signal $E_{drop}$ at the drop port. An external optical feedback link connects the through and the add port with time delay $\tau_{F}$, under specific feedback strength $\eta_{F}$ and phase ($\phi_{F}$) conditions.

The temporal dynamics of the MRR is commonly described, within the coupled-mode theory framework, by the following set of coupled differential equations \cite{Johnson2006}:

\begin{align}
    \frac{dU(t)}{dt} &= \bigg[ -i(\omega_p-\omega_{o}(t)) - \gamma(t)\bigg] U(t) + i\sqrt{2\Gamma}(E_{inp}(t)+E_{add}(t)).
    \label{eq_Uint}
    \\
    \frac{d\Delta N(t)}{dt}	&= -\frac{\Delta N(t)}{\tau_{FC}}+G_{TPA}\left|U(t)\right|^4.
    \label{eq_N}
    \\
    \frac{d\Delta T(t)}{dt}	&= -\frac{\Delta T(t)}{\tau_{TH}}+\frac{P_{abs}}{m c_p}.
    \label{eq_T}
\end{align}

\noindent Eq. (\ref{eq_Uint}) describes the variation of the optical energy amplitude $U(t)$ within the MRR, when considering an input electrical field $E_{inp}$ and a feedback field $E_{add}$, at the same frequency $\omega_p=2\pi c/\lambda_p$. Eq. (\ref{eq_N}) describes the excess of the free carrier density ($\Delta N$) within the MRR, that is generated by TPA, at a rate of $G_{TPA}$, and is recombined with a decay constant $\tau_{FC}$. Both the creation and the recombination of free carriers are assisted by the emission of phonons in the silicon waveguide, with a consequent variation $\Delta T$ of the mode-averaged temperature. This is described as Newton's law in Eq. (\ref{eq_T}), where $P_{abs}$ is the absorbed power by the material that causes heating, $m$ indicates the mass of the ring, $c_p$ is the silicon specific heat and $\tau_{TH}$ indicates the thermal decay time due to the heat dissipation with the surrounding medium.

Thermal and free carrier changes modify the refractive index of the MRR, introducing nonlinear effects, as it emerges from Eq. (\ref{eq_Uint}) through the terms $\omega_{o}(t)$ and $\gamma(t)$. 
The first term $\omega_{o}(t)$ is described by $\omega_{o}(t)=\omega_{o}+\delta \omega_{nl}(t)$, with $\omega_{o}$ being the resonance frequency in absence of nonlinearities and $\delta \omega_{nl}(t)$ the nonlinear contribution:

\begin{equation}
    \delta \omega_{nl}(t)= \frac{\Gamma_c}{n_{Si}}\bigg( \frac{dn_{Si}}{dT} \Delta T(t) + \frac{dn_{Si}}{dN} \Delta N(t) \bigg).
    \label{eq_ResShift}
\end{equation}

\noindent In Eq. (\ref{eq_ResShift}), $\Gamma_c$ is the modal confinement factor and $n_{Si}$ the Silicon refractive index. 
The second term $\gamma(t)$ includes the losses that are present in the linear operation $\gamma_{lin}$ and the losses induced by TPA and free carrier absorption (FCA):

\begin{equation}
    \gamma(t)=\gamma_{lin}+\eta_{FCA} \Delta N(t)+ \eta_{TPA}\left|U(t)\right|^2,
\end{equation}

\noindent where $\gamma_{lin}=\gamma_{i}+2\gamma_{e}$ accounts for the MRR intrinsic losses rate $\gamma_{i}$ (due to material absorption, scattering, bending) and the extrinsic losses rate $2\gamma_{e}$ (due to the symmetric coupling with the two straight waveguides). The characteristic timescale of a MRR in linear regime is related to the photon lifetime of the cavity $\tau_{ph}=\gamma_{lin}^{-1}$, when $\delta \omega_{nl}(t)=0$ and $\gamma(t)=\gamma_{lin}$, and represents the decay rate of the MRR's internal optical power. In a nonlinear regime, activated by the TPA, $\tau_{FC}$ and $\tau_{TH}$ become also important for the dynamics. In a rough approximation, $\tau_{FC}\approx 10^{-2} \tau_{TH}$. Thus, depending on the temporal scale of the input signal, the MRR can exhibit dynamics that are influenced by only one of these nonlinear effects. In the current work, we focus on the nonlinear effects triggered by the free carriers excited in the MRR waveguide, as we aim to process information that is encoded at this time scale.\\
We compute the electric field using a scattering matrix approach at the through and drop ports of the MRR ($E_{th}$, $E_{drop}$), while at the add port we feed the delayed signal from the through port ($E_{add}$), after tuning its amplitude and phase:

\begin{align}
    E_{th}(t) = t_{r} E_{inp}(t) + \sqrt{2\gamma_{e}} U(t).
    \label{eq_Eth}
    \\
    E_{drop}(t) = \sqrt{2\gamma_{e}} U(t) + t_{r} E_{add}(t).
    \label{eq_Edrop}
    \\
    E_{add}(t) = \sqrt{\eta_{F}}e^{-i\phi_{F}} E_{th}(t-\tau_{F}).
    \label{eq_Eadd}
\end{align}

\noindent $t_{r}$ indicates the field amplitude transmission from the input (add) port to the through (drop) port (see Appendix). In Eq. (\ref{eq_Eadd}), $\tau_{F}$ is the delay introduced by the feedback line, while $\eta_{F}$ and $\phi_{F}$ modify the through signal before re-entering the MRR. $\eta_{F}$ varies in the range [0 , 1], with 0 representing a completely attenuated signal and 1 being a lossless re-injection of the through signal in the add port. The coupling phase condition $\phi_{F}$ is expressed as: 

\begin{equation}
    \phi_{F}=\beta_{F}*L_{F} +\Delta\phi_{F},
    \label{eq_phi_F}
\end{equation}

\noindent where $\beta_{F}=\frac{2\pi n_{F}}{\lambda_{p}}$ is the propagation constant along the feedback line, $n_F$ is the refractive index of the feedback line, and $\Delta\phi_{F}$ is an external phase shift that ranges in [0 - 2$\pi$]. Dispersion and nonlinear effects on light propagating along the feedback line are not here considered. For the dynamical investigation of the system, we define a starting wavelength detuning between the laser wavelength and the MRR resonance equal to: $\Delta\lambda_{s}=\lambda_p-\lambda_{o}$, and a resonance shift induced by nonlinear effects $\Delta\lambda_{o}(t)=\lambda_{o}(t)-\lambda_{o}$, where $\lambda_{o}(t)=2\pi c/w_{o}(t)$ and $\lambda_{o}=2\pi c/w_{o}$. The set of Eq. (\ref{eq_Uint})-(\ref{eq_T}) are numerically integrated using a Runge–Kutta method, with an integration step of 2 ps, which is sufficient to account for the lowest timescale effects ($\tau_{ph}\approx$ 50 ps). The model considers only a unidirectional propagation and the values of the parameters are reported in Table \ref{tab:parameters}, in the appendix.

\label{sec_Model}

\begin{figure}[b]
\centering\includegraphics[scale=0.22]{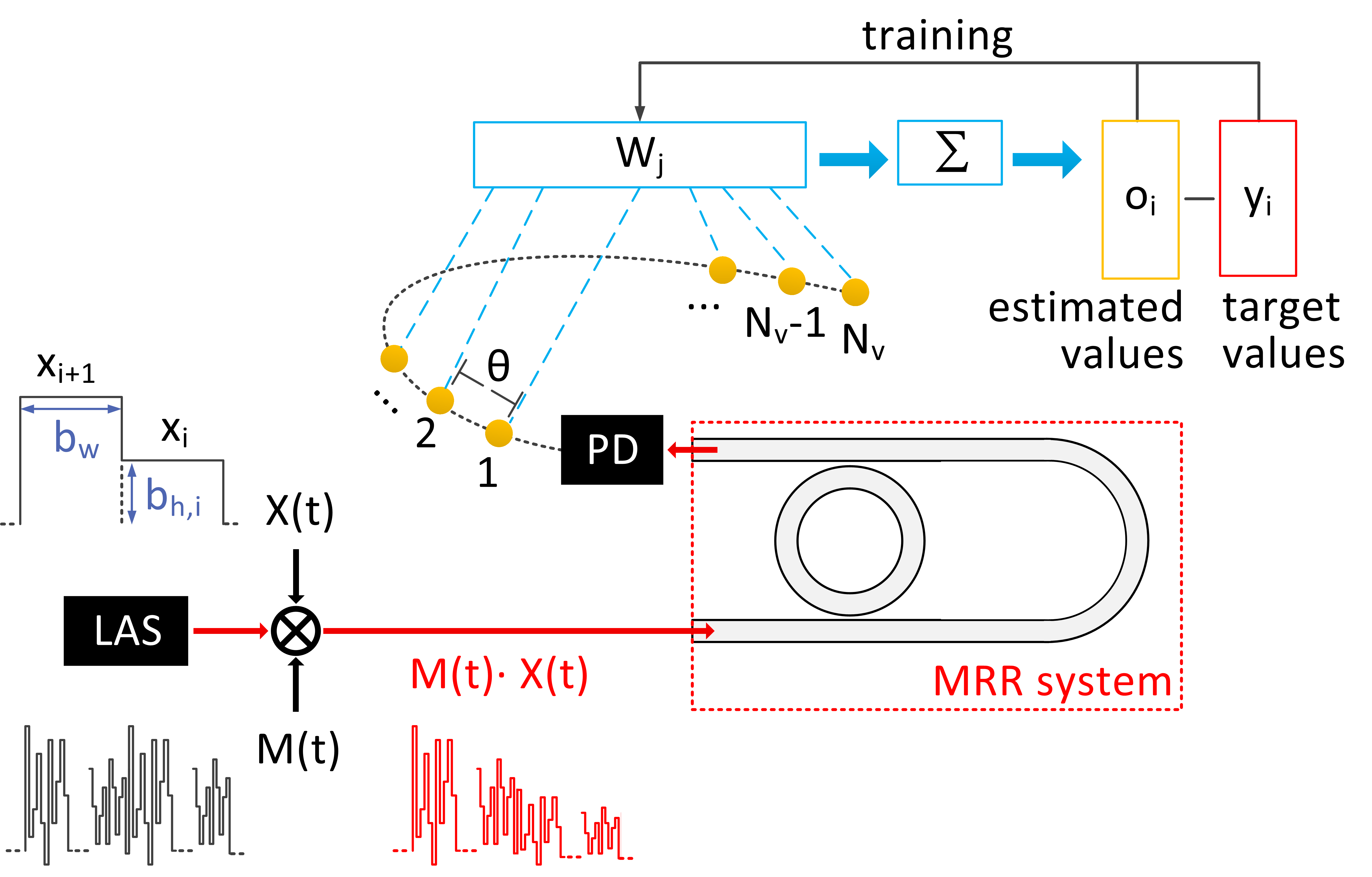}
\caption{Schematic of time delay RC with an MRR subject to optical feedback. The encoded information $X(t)$ is masked with a sequence $M(t)$ and modulates the optical power from the laser (LAS) emission. At the drop port, the photodetected (PD) signal provides the time-multiplexed output states of the reservoir, which are weighted and linearly combined to compute the predicted value $o_i$. The weight optimization is performed via a linear classifier, with supervised learning over the expected values $y_{i}$ data set.}
\label{System_scheme}
\end{figure}

\section{MRR in time delay RC}
The scheme of our time-delay RC follows the typical formulation, consisting of an input layer, the reservoir, and an output layer (Fig. \ref{System_scheme}). At the input layer, we introduce into the system the information to be processed. As discussed in \cite{Appeltant_2011}, the input information can be continuous in time or with discrete values. In both cases, sampled values $x_i$ from the sequence are extracted and codified in bit of duration $b_w$ and amplitude $b_{h,i}$ ($X(t)$). Every bit is masked with a set of random values $M(t)$ taken from a uniform distribution. The size of the mask set $M(t)$ is equal to the number or virtual nodes $N_{v}$ that are defined in the reservoir \cite{Appeltant_2011} and is periodic, with period $b_w$. Thus, $M(t)=M(t+b_w)$, with an intra-mask temporal distance between its values equal to $\theta$, with $\theta$ being lower than all the characteristic times of the MRR. The resulting signal modulates the optical carrier emitted by a laser, with maximum optical power $P_{max}$, and at an emission wavelength detuned by $\Delta\lambda_{s}$ with respect to the initial MRR wavelength resonance. Then, the optical signal enters the input port of the MRR system and propagates both along the MRR and the feedback line. The received optical signal at the drop port is photodetected and synchronously sampled at the masking sampling distance $\theta$. We also include noise in the optical system, with $40$dB signal to noise ratio for the $P_{max}$ operation. The nonlinear transformation of the input signal occurs at both the MRR system and the photodetection stage. Eventually, each uni-dimensional input information is projected through this physical system into a higher dimensional space defined by the number of virtual nodes $N_{v}=\frac{b_w}{\theta}$. By defining $N_{j,i}$ as the $j^{th}$ virtual node response associated with the $i^{th}$ processed element $x_i$, we represent numerically the above operation, by considering the corresponding electric field of the optical signal:

\begin{align}
    E_{inp}(t) &= [X(t) M(t)]^{1/2} =  [x_i m_j]^{1/2} \text{\hspace{0.2em},  for  } b_w(i-1)+\theta(j-1)\leq t\leq b_w(i-1)+\theta j,
    \label{eq_s_&_h}
    \\
    N_{j,i} &= \mid{E_{drop}(b_w(i-1)+\theta j)}\mid^2,
    \label{eq_VNmatrix}
\end{align}

\noindent where $m_j$ indicates the $j^{th}$ mask value with $j=1,\dots, N_{v}$ and $i$ indicates the index of the input element that is processed. Finally, at the output layer, we obtain a unique value estimator $o_i$, related to the input $x_{i}$, given as a linear combination of the corresponding virtual nodes' responses:

\begin{equation}
    o_i = \sum_{j=1}^{N_{v}} W_{j} N_{j,i},
    \label{eq_outLayer}
\end{equation}

\noindent where $W_{j}$ is a $N_{v}$-dimensional vector of the readout weights, which are trained via a linear regression algorithm to minimize the normalized mean square error (NMSE) between $o_i$ and an expected value $y_{i}$:

\begin{equation}
    NMSE=\frac{\sum_{i} (o_i-y_{i})^2}{N_d \sigma_y^2}.
    \label{eq_NMSE}
\end{equation}

\noindent In Eq. (\ref{eq_NMSE}), the sum includes all the elements of the dataset $N_d$. These weights are then used to evaluate the operation of the system on independent testing datasets. The lower the $NMSE$ is, the better the system predicts the expected output series. 

\label{Sec_RCscheme}
    
\subsection{Characteristic timescales and nonlinearity}

\begin{figure}[t]
\centering\includegraphics[scale=0.5]{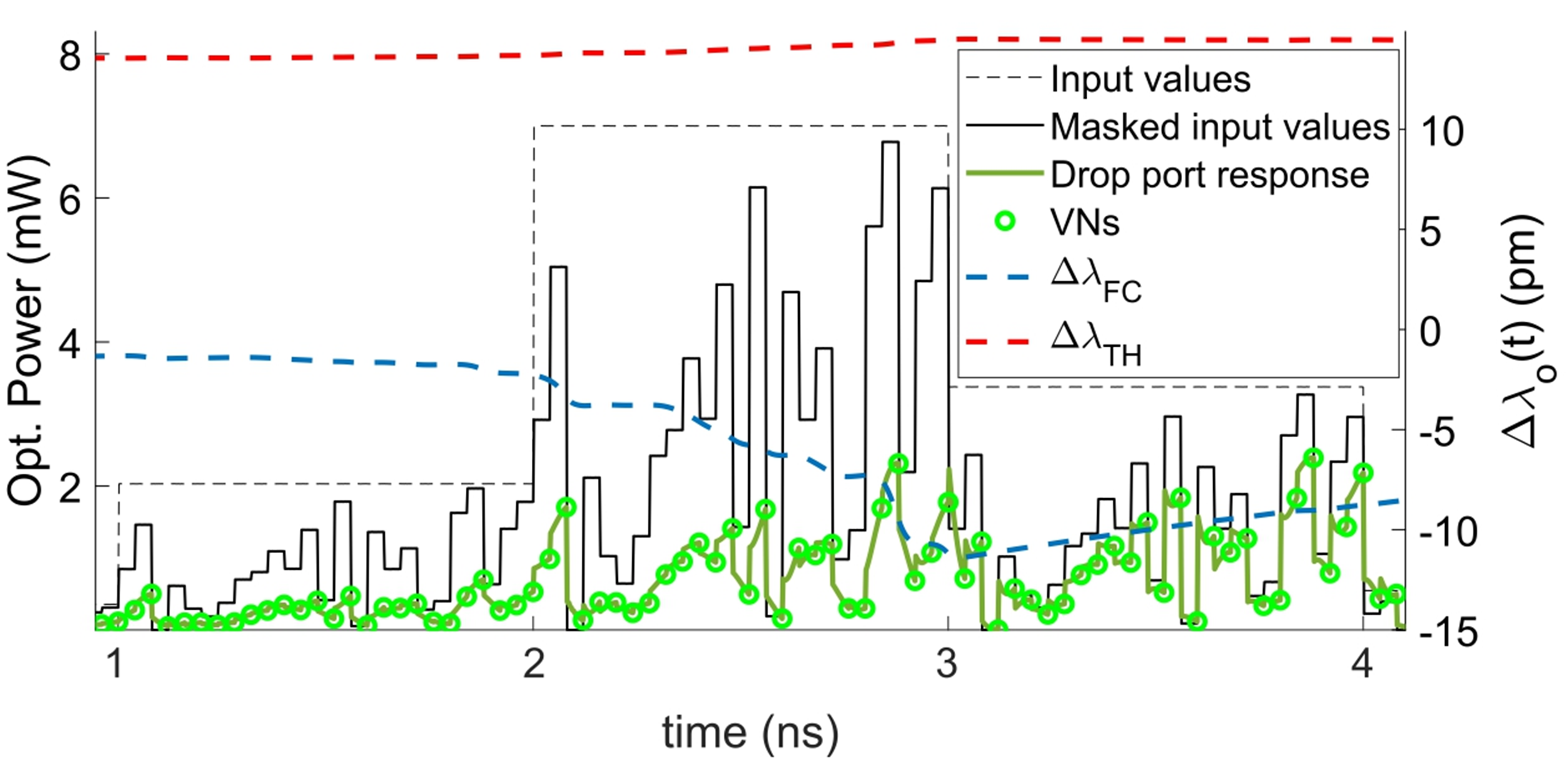}
\caption{Dynamical response of an MRR to a fast modulating signal at the input port, based on the input layer concept of a time delay RC scheme. $X(t)$ (black dashed line) and $X(t)M(t)$ (black continuous line) represent the input information before and after the masking, respectively. The response of the MRR is obtained at the drop port (green line), while the green circles indicate the sampled response of the virtual nodes of the reservoir, separated in time by $\theta=40$ ps (one virtual node per mask node). The blue and red dashed lines show the contribution of the free carrier ($\Delta\lambda_{FC}$) and the thermal ($\Delta\lambda_{TH}$) nonlinear effects, respectively, to the wavelength resonance shift (shown in the right y-axis).}
\label{Setting_Example}
\end{figure}

The MRR has a quality factor of $Q=3.19\times 10^4$ and can exhibit self-pulsation dynamics, a phenomenon that relies on a free carrier concentration variation in the waveguide \cite{Vaerenbergh2012}. As discussed in section \ref{sec_Model}, three different timescales characterize the MRR operation: the photon lifetime (here $\tau_{ph}\approx 50$ ps), the free carrier lifetime (here $\tau_{FC} \approx 3$ns) and the thermal lifetime (here $\tau_{TH} \approx 83$ ns), with the last two being associated with the MRR nonlinearity. Here the masking samples are applied every $\theta=40$ ps < $\tau_{ph}$; in this way, the fastest characteristic time response of the MRR $\tau_{ph}$ - associated with the photon lifetime - is also exploited to keep the system's operation in a transient state. The MRR does not completely discharge the internal field, when the next mask sample is applied. This allows to retain information of previous states (short memory) and in this way neighboring virtual nodes are coupled through inertia. In our approach, we exploit the free carrier nonlinearity, which has a faster time response and allows for faster computations. Here we select an information encoding duration of $b_w = 1$ ns $\approx \tau_{FC}$. This allows each encoded input bit $x_i$ to trigger a measurable change of the free carriers inside the MRR - in presence of high values of $b_{h,i}$. Consequently, $N_{v} = 25$ virtual nodes are defined within a duration of $b_w$. This number of virtual nodes allows computations at GHz rates and is also compatible with delays provided by integrated silicon feedback waveguides \cite{Harkhoe2020}. In Fig. \ref{Setting_Example} we show the response of neighboring virtual nodes (green line), strongly coupled to the previous states. We also show the contribution of the free carriers (blue dashed line) and the thermal effects (red dashed line) to the resonance shift. The optical input after masking (black line) enters into the MRR and drives the generation of the free carriers that shift the resonance by $\Delta\lambda_{FC}$. At the same time, thermal effects cause a shift by $\Delta\lambda_{TH}$, but they are too slow compared with the input changes, so that they do not contribute to the nonlinear transformation of the optical signal and only add a positive bias to the resonance position. Thermal effects become important - particularly for high quality-factor MRRs - when the optical power within the MRR is high enough to activate self pulsations. By adopting smaller information encoding durations (e.g. $b_w < 1 ns$) we further limit the number of virtual nodes and the computational power of the reservoir becomes limited. On the other hand, by adopting larger information encoding durations (e.g. $b_w > 1 ns$) we are able to introduce a larger number of virtual nodes and improve the performance of some computational tasks (see section 5) but by reducing the computational speed.

\label{sec_setting}

\section{Results}
\label{Sec_results}
Under the above conditions, the MRR with optical feedback system is tested on three different benchmark computational tasks, that have different requirements for signal processing. The NARMA 10 belongs to the category of nonlinear system identification tasks and requires explicitly 10 memory steps to be solved. The Mackey-Glass and the Santa Fe are benchmark one-step-ahead chaotic time-series prediction tasks where the system has to predict a future value $x_{i+1}$ of the input series, while processing $x_i$. 

Since the performance of every task relies on a characteristic trade-off between a nonlinear transformation of the input information and the system's linear memory \cite{NLvsLMemory2017}, an estimation of these two quantities is addressed below. The former is evaluated indirectly, via the standard deviation of the wavelength resonance shift $\sigma(\lambda_{o}(t))$. A higher standard deviation indicates higher MRR nonlinearity. The latter is evaluated by the linear memory capacity ($MC$) task, as we present it here. $MC$ is traditionally calculated by using as an input series a random sequence of bits, with values taken from a uniform distribution \cite{Jaeger_MC}. However, there is an inconsistency in this method when one tries to evaluate the response of a nonlinear system to an input with specific spectral properties. The linear memory capacity of the system can be different when entering into the system either a random sequence or a sequence with correlated temporal profiles. This stands also in our case, where different input series may lead the MRR operation under different nonlinear dynamics, even for the same operating parameters. For this reason, in this work, the $MC$ task is solved for the actual timeseries of the benchmark tasks we will evaluate. The system is trained to remember the $l^{th}$ previous input element of the used series, by exploiting the information that is still present in the system. It is computed as:

\begin{equation}
    MC=\sum_{l=0}^{l_{max}} m(l), \quad\text{with}\quad m(l)= \frac{cov^2(o(n), x(n-l))}{\sigma_{o}^2 \sigma_{x}^2}.
    \label{eq_MC}
\end{equation}

\noindent $m(l)$ measures the normalized linear correlation between the predicted ($o(n)$) and delayed ($x(n-l)$) input series, with $cov^2()$ indicating the covariance between two vectors and $\sigma^2$ the variance. When this correlation is very small ($<<1$), the system is unable to preserve any information of $l$ past input. On the contrary, when $m(l)$ approximates one, the system remembers the exact value. For the MC computation, we also consider the case of $l = 0$ that refers to the capability of the system to retrieve the actual input.

We investigate the performance of the MRR system on the selected tasks, for different configurations, by tuning the critical operational parameters, such as the starting wavelength detuning $\Delta\lambda_{s}$, the maximum input optical power $P_{max}$, the feedback phase $\Delta\phi_{F}$ and the feedback strength $\eta_{F}$. All these parameters affect the optical power circulating within the MRR and thus its nonlinear operation. We select the following range values for these parameters: $\Delta\lambda_{s} \in [-50$ pm, $50$ pm], with step of $10$ pm, so that all the MRR resonance (having Full Width at Half Maximum $FWHM=48$pm) is covered; $P_{max} \in [1$ mW, $8$ mW], with step of $1$ mW, including also the value $0.1$ mW where the MRR operates in a linear regime; $\Delta\phi_{F} \in [0, 2\pi]$; $\eta_{F} \in [0, 1$]. For training the MRR system to the different tasks, we use 1000 input data values to drive the system in a working regime and eliminate any oscillatory operation due to the inclusion of the input. Then we use the next 2000 input data values for training and the next 1000 data for testing the system on previously unseen entries. The same random mask $M(t)$ is used in all the simulations that involve the same number of virtual nodes, which is fixed to 25 unless differently specified. The ridge regression parameter of the RC's output layer linear classifier is set to $10^{-4}$. 

\label{results_intro}
\subsection{Narma 10 benchmark test}

\begin{figure}[t]
\centering\includegraphics[scale=0.13]{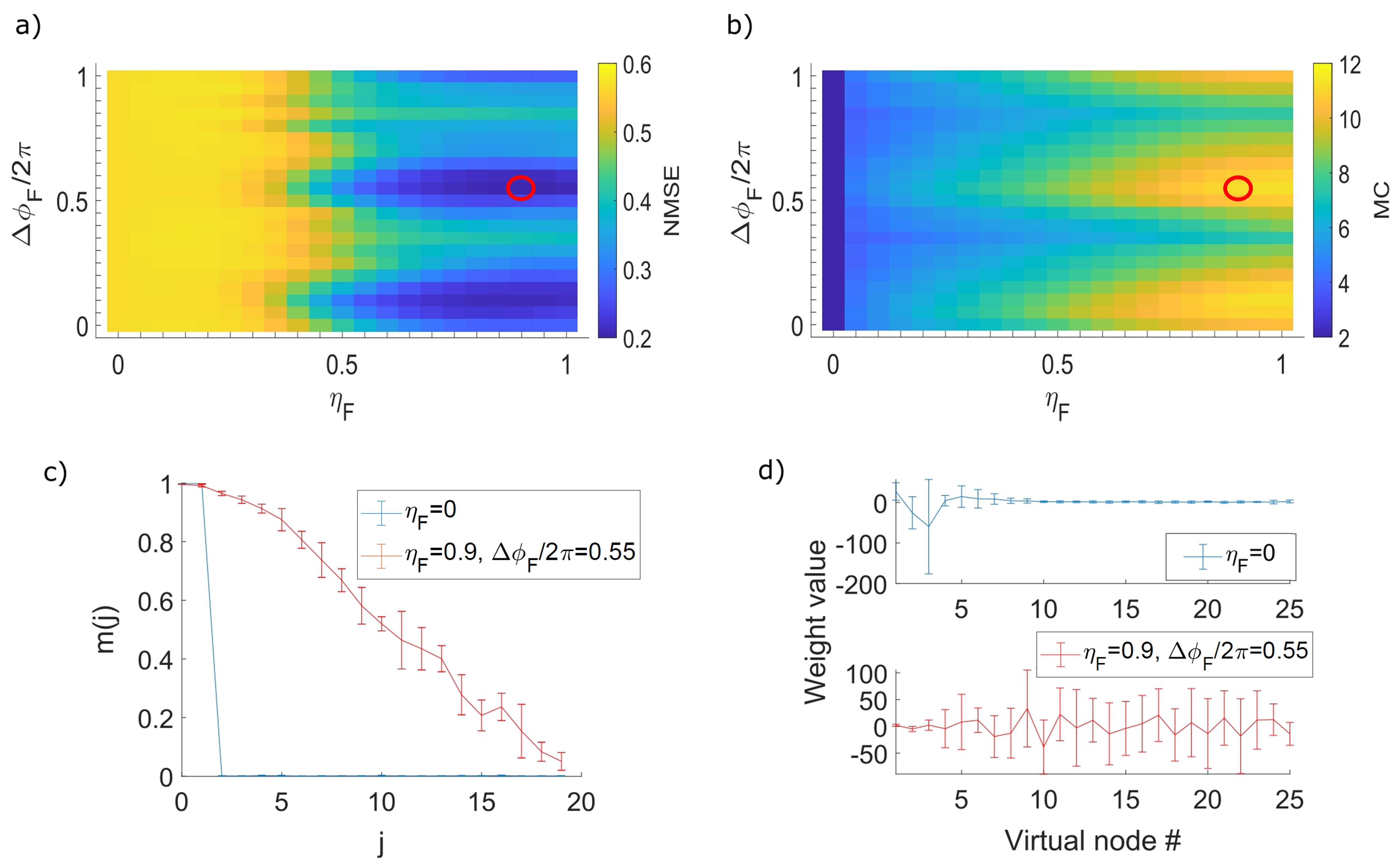}
\caption{Performance of the Narma 10 benchmark task. (a) NMSE and (b) MC, versus optical feedback strength $\eta_{F}$ and phase $\Delta\phi_{F}$. Red circle denotes the conditions with the lowest NMSE. (c) Memory function $m(l)$, for the cases without feedback (blue line) and with feedback conditions that result in the lowest NMSE (red line). (d) Readout weights for a task to remember the previous input value $x_{i-1}$, for the cases without feedback (blue line) and with feedback conditions that result in the lowest NMSE (red line). MC is computed using $l_{max} = 19$. The initial wavelength shift is $\Delta\lambda_{s} = -10 pm$ and the MRR is operating in the linear regime, with $b_w = 1$ ns.}
\label{NT_25VNs}
\end{figure}

In the Narma 10 task, our system is trained to predict the response of a discrete-time tenth order nonlinear auto-regressive moving average (NARMA) system \cite{Atiya2000}, described by:

\begin{equation}
    r_{i+1} =  0.3r_{i} + 0.05 r_{i} (\sum_{j=0}^{9} r_{i-j}) + 1.5 x_{i-9} x_{i} + 0.1, 
    \label{eq_NT}
\end{equation}

\noindent where $x_{i}$ represents the $i^{th}$ element of the input series uniformly distributed in the range [0 , 0.5] and $r_{i+1}$ is the correspondent expected target ($y_{i}$ in Fig. \ref{Sec_RCscheme}). This task requires explicitly at least 10 values (the current one and 9 in the past) to be considered to predict the next value. In Fig. \ref{NT_25VNs}(a) we show the NMSE performance of the MRR system, versus the feedback parameters $\eta_{F}$ and $\Delta\phi_{F}$. In parallel, we show the linear MC of the MRR system for this task in Fig. \ref{NT_25VNs}(b). While for the performance optimization we use the 4-dimensional parameter manifold ($P_{max}$, $\Delta\lambda_{s}$, $\eta_{F}$, $\Delta\phi_{F}$), we provide our results in two dimensions, while fixing the rest of the parameters of the complete manifold.
We find that the parameter space for which we observe the lowest values of NMSE is where the MC approaches its maximum value. In absence of the external optical feedback ($\eta_{F}=0$), the MC value is only around 2. This very limited memory emerges from the single MRR operation and the inertia between the last virtual nodes' responses of the input value $x_{i-1}$ and the first virtual nodes' responses of the next input value $x_{i}$. In this case, the system can remember the input $x_{i-1}$ from the actual input $x_{i}$ (Fig. \ref{NT_25VNs}(c), blue line). This can be verified by the computed weight values of the RC linear classifier, as shown in Fig. \ref{NT_25VNs}(d). When we train our classifier to provide as an output the previous value of the series, by considering the response of the reservoir to the actual input $x_{i}$, only the response of the first virtual nodes is important for computation (Fig. \ref{NT_25VNs}(d), blue line). But when activating the feedback (Fig. \ref{NT_25VNs}(c), red line), all virtual nodes contribute to the task computation (Fig. \ref{NT_25VNs}(d), red line). To obtain an extended linear memory, a strong feedback parameter $\eta_{F}$ is required, under an appropriate phase condition at the add port of the MRR. 

For the NARMA 10 task, the minimum error $NMSE_{min}= 0.204\pm 0.026$ is found at $\eta_{F} = 0.9$, and $\Delta\phi_{F}/2\pi = 0.55$, for $P_{max} = 0.1 mW$ and a starting wavelength detuning of $\Delta\lambda_{s} = -10 pm$. Thus, the MRR is operating in a linear regime. The combination of a linear MRR with an optical feedback delay, acts like a linear analog shift register. If the MRR is in resonance, part of the feedback signal is coupled back to the MRR. Thus, an input light pulse can propagate multiple times through the MRR system, providing a linear optical memory to the system. According to \cite{NLvsLMemory2017}, a worse performance is expected when the MRR operates in a nonlinear regime, since it progressively distorts the information. The lowest obtained NMSE value is even higher than the one expected from a linear shift register ($NMSE_{SR} = 0.16$) \cite{Appeltant_2011}. The reason for which we get a higher NMSE is due to the small number of virtual nodes. But since in this task the MRR dynamics is not bounded to its nonlinearity, a longer bit duration $b_{w}$ can be considered while preserving the same dynamical response and the same virtual node time separation $\theta=40$ps. For example, when $N_{v} = 200$ and $b_w = 8$ ns, we obtain $NMSE = 0.010\pm 0.009$, an equivalent performance with the one reported in \cite{Appeltant_2011}. This improvement, compared to the linear shift register performance, is attributed to the square-law nonlinearity of the photodetection, since both cases exploit the same linear memory. This operation is consistent with the one presented in \cite{Vinckier2015}, where a linear external cavity with an optical fiber loop was used and from which the output optical signal was photodetected. For comparison, the use of the MRR in absence of optical feedback, results in $NMSE= 0.545\pm 0.001$, for $P_{max}=2$ mW and $\Delta\lambda_{s}=20$ pm.

In conclusion, the single MRR operating in the linear regime without feedback can preserve the previous value of input information, through inertia, at MRR photon lifetime time scales. This memory, along with the photodetection nonlinearity, is sufficient to solve one-step-before memory tasks. For the Narma 10 task, which has longer memory requirements, the external cavity is the main contributor to the linear memory capacity of the computing system.

\label{sec_NT}

\subsection{Mackey-Glass benchmark test}
The Mackey-Glass input series is obtained by integrating in time the following equation:  
\begin{equation}
    \frac{dx(t)}{dt}=\frac{\alpha x(t-\tau)}{1+x(t-\tau)^\beta}-\gamma x(t).
    \label{eq_MG}
\end{equation}

\noindent Eq. (\ref{eq_MG}) can provide a rich variety of periodic, aperiodic, and chaotic solutions. It was initially used in \cite{MG1977} to describe physiological diseases in the human body and later, in recurrent neural networks \cite{ESN_2004}, as a benchmark timeseries for prediction. In the last case, a weakly chaotic behaviour is obtained, by numerically solving Eq. (\ref{eq_MG}) with an integration step of $0.1$, and the following parameter values: $\alpha = 0.2$, $\beta = 10$, $\gamma = 0.1$, and $\tau = 17$. In our investigation, we apply an oversampling of $3$, similarly to \cite{Bueno2017}.

\begin{figure}[t]
\centering\includegraphics[scale=0.65]{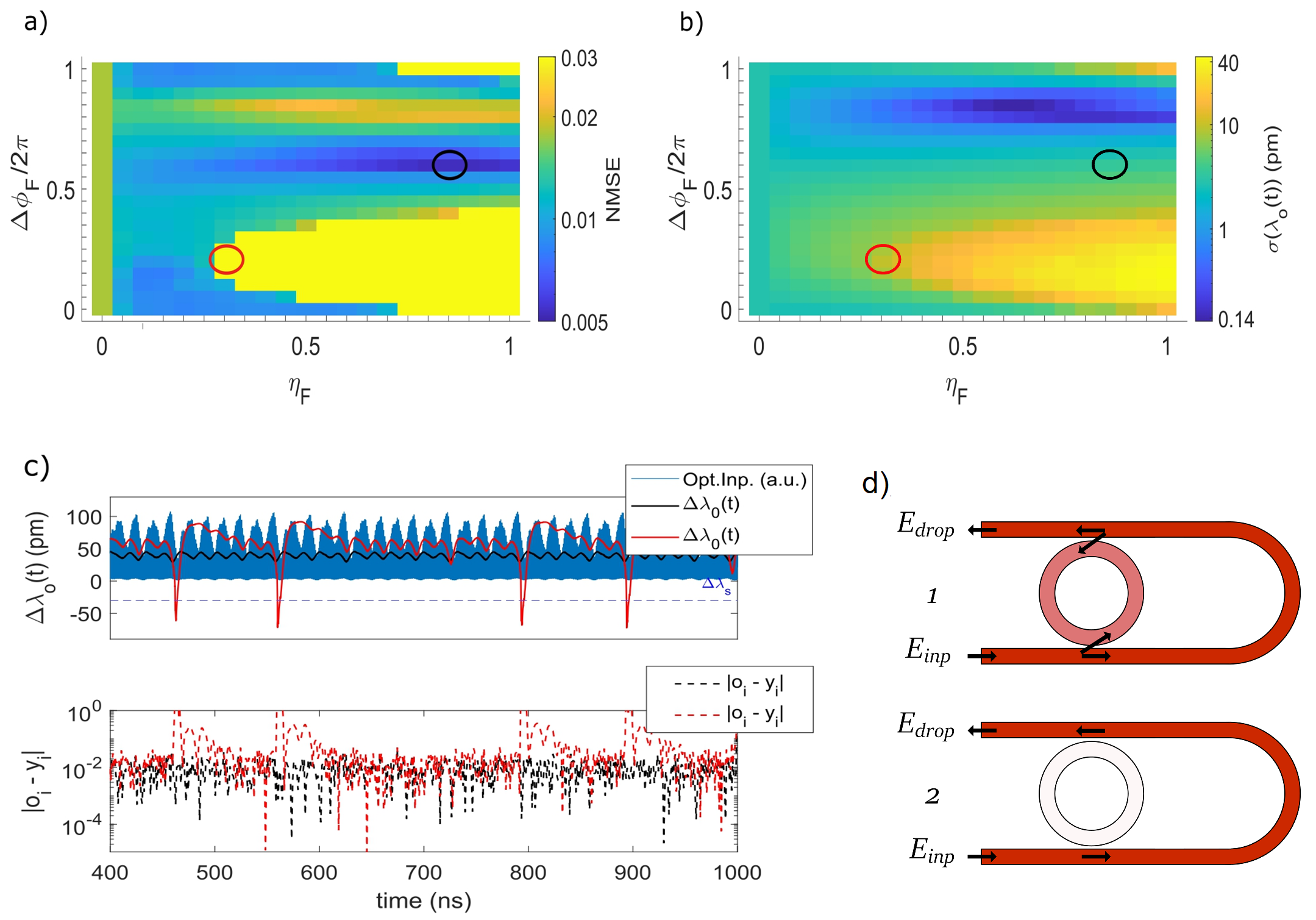}
\caption{Performance of the Mackey-Glass benchmark task. (a) NMSE and (b) standard deviation of the resonance wavelength shift $\sigma (\lambda_0)$, versus optical feedback strength $\eta_{F}$ and phase $\Delta\phi_{F}$ of the MRR system. Black (red) circle denotes the conditions with the lowest (highest) NMSE. (c) Temporal evolution of the resonance shift and the bit error $|o_i - y_i|$ during the task for two feedback conditions: the black line corresponds to the lowest NMSE (black circle, (a)), and the red line corresponds to the highest NMSE (red circle, (a)). (d) Dynamical operation of the MRR with optical feedback under self pulsations: light occasionally enters (path 1, upper) or bypasses (path 2, lower) the MRR. The initial wavelength shift is $\Delta\lambda_{s} = -30 pm$, the maximum launched optical power at the input is $P_{max} = 5$ mW and $b_w = 1$ ns.}
\label{MG_25VNs_Lnoise}
\end{figure}

The lowest $NMSE$ value for this prediction task is obtained for $P_{max} = 5$ mW, $\Delta\lambda_{s} = -30$ pm and for feedback conditions that are shown in Fig. \ref{MG_25VNs_Lnoise}(a) - black circle ($\eta_{F} = 0.85$ and $\Delta\phi_{F}/2\pi = 0.6$). These conditions result in an $NMSE_{min} = 0.0053\pm 0.0005$, lower than $NMSE_{SR} = 0.01$ obtained by a linear shift register. Differently from the NARMA 10 task, the optimal computing conditions exploit the MRR nonlinearity, as indicated by the corresponding standard deviation value of the MRR's resonance shift $\Delta\lambda_{o}(t)$ (Fig. \ref{MG_25VNs_Lnoise}(b)). Nevertheless, the resonance shift must be constrained within some boundaries, so that the MRR does not get out of resonance. The external cavity, besides its contribution to the extended fading memory, is beneficial in the following sense: different $\eta_{F}$ and $\Delta\phi_{F}$ values result in different interference conditions between the feedback signal and the internal field of the MRR. Thus, the feedback conditions also determine the circulating internal optical power in the MRR and eventually its nonlinearity. For example, when $\eta_{F} > 0.3$ and $0 < \Delta\phi_{F}/2\pi < 0.45$, a constructive interference is observed in the MRR, leading to higher values of $\Delta\lambda_{o}(t)$, thus higher MRR nonlinearity (Fig. \ref{MG_25VNs_Lnoise}(b)) and degraded performance (Fig. \ref{MG_25VNs_Lnoise}(a)). Under this configuration, the MRR system is driven out of resonance. This is illustrated in Fig. \ref{MG_25VNs_Lnoise}(c, upper panel, red line) from the evolution of the resonance shift $\Delta\lambda_{o}(t)$ in time, while executing the computation. For the conditions indicated with the red circle in Fig. \ref{MG_25VNs_Lnoise}(a), a series of bursting spikes in the wavelength resonance shift is observed, followed by a thermal warming and then a thermal cool-down. This dynamical behavior resembles self pulsations, where the oscillation occurs whenever the resonance becomes too detuned with respect to the pumping wavelength (out of resonance condition). In these time intervals, light mainly propagates through the external cavity (Fig. \ref{MG_25VNs_Lnoise}(d, path 2)) and is not susceptible anymore to the MRR nonlinearity. At the same time, as the feedback signal bypasses the MRR, it can not be coupled back and iterate further. In this way the system loses also the feedback memory. These conditions result in degraded performance, as it is indicated by the corresponding higher prediction error in Fig. \ref{MG_25VNs_Lnoise}(c, bottom panel, red line). As a comparison, the lowest NMSE configuration is also reported in Fig. \ref{MG_25VNs_Lnoise}(c, black line). In this case $\Delta\lambda_{o}(t)$ oscillates in phase with the input optical peaks (Fig. \ref{MG_25VNs_Lnoise}(c, upper panel, blue line)) and with lower amplitudes. In a comparison with the NARMA 10 task, we observe that the region with worst NMSE performance (Fig. \ref{MG_25VNs_Lnoise}(a)) – where the MRR transition to self-sustained oscillations occurs, due to the competition between thermal and free carrier nonlinearities – differs from the one in Fig. \ref{NT_25VNs}(a). In the latter, the linear MRR operation does not induce self-sustained oscillations. In a comparison with the linear shift register, neither the single MRR in absence of the feedback ($NMSE = 0.015 \pm 0.002$), nor the linear MRR in presence of feedback ($NMSE = 0.0095 \pm 0.0009$) provide an improved performance. This can be only obtained by combining the MRR nonlinearities with the memory provided by the external cavity.

\subsection{Santa Fe benchmark test}
The Santa Fe benchmark test is the second one-step-ahead time series prediction task we investigate. In this task, the input series is the optical power emitted by a far-infrared laser that operates in a chaotic regime \cite{SF93}. This publicly available dataset has experimental noise in its values, in contrast to the Narma 10 and the Mackey-Glass timeseries.  

\begin{figure}[htbp]
\centering\includegraphics[scale=0.35]{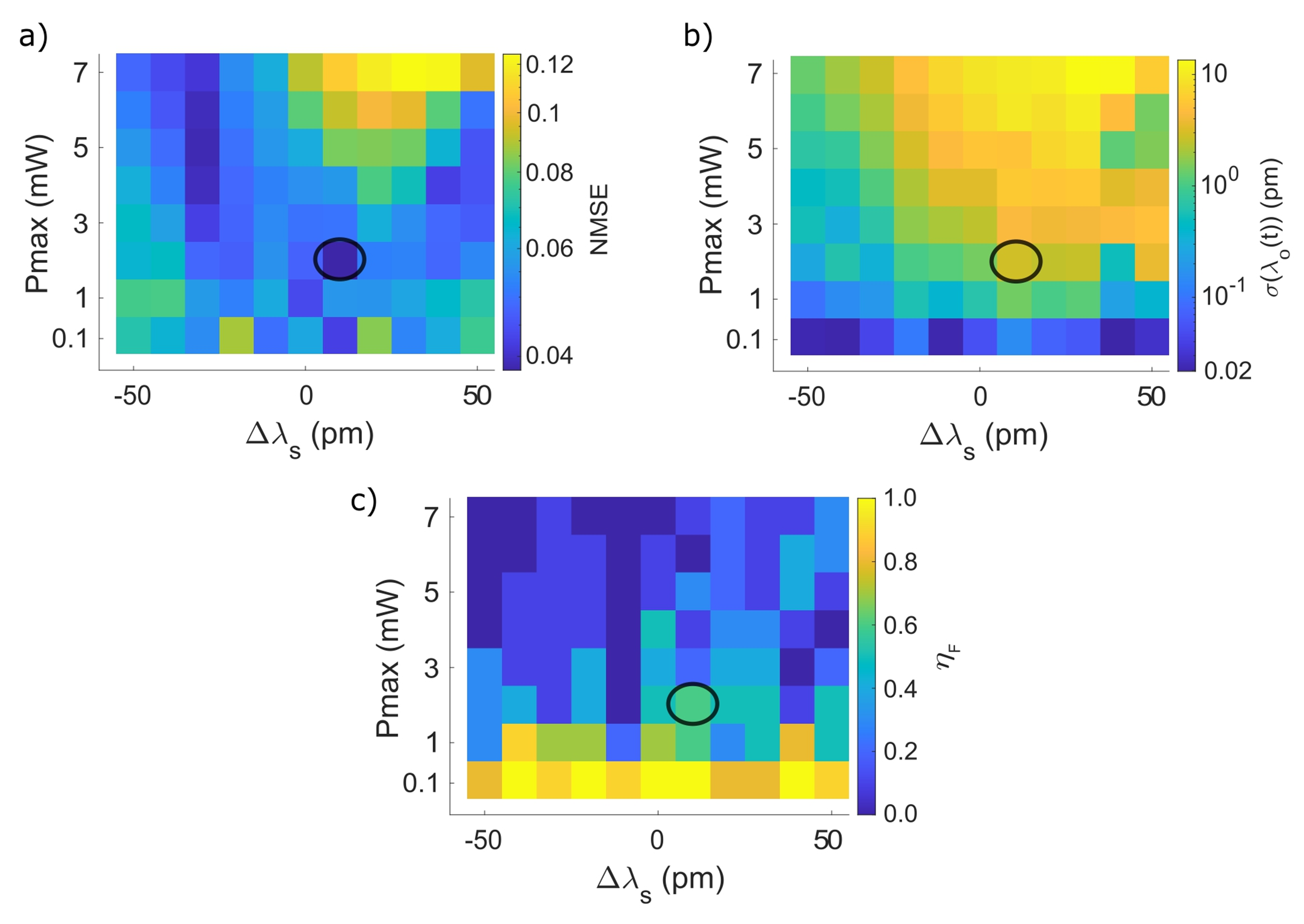}
\caption{Performance of the Santa Fe benchmark task, by using a single MRR with external feedback. (a) NMSE, (b) standard deviation of the resonance wavelength shift $\sigma (\lambda_0)$, and (c) strength of the feedback, versus the starting detuning $\Delta\lambda_{s}$ and the maximum incident power $P_{max}$. The quantities refer to the feedback configuration with the lower error achieved at each ($\Delta\lambda_{s}$, $P_{max}$). The black circle indicates the conditions for which we obtain the lowest NMSE.}
\label{SF_WithFeed_general}
\end{figure}

Processing this dataset with a linear shift register results in a value of $NMSE_{SR} = 0.2$. By considering MRR-based processing, this error is significantly reduced. This is shown in Fig. \ref{SF_WithFeed_general}(a), where we plot the best NMSE for each pair ($\Delta\lambda_{s}$, $P_{max}$) investigated. We find an NMSE as low as $0.038\pm0.008$, for the following parameters’ configuration: $\Delta\lambda_{s}=10$ pm, $P_{max} = 2$ mW, $\eta_{F}=0.55$ and $\Delta\phi_{F}=0$ ((Fig. \ref{SF_WithFeed_general}(a), black circle, with feedback parameters not displayed). We also monitor the standard deviation of the resonance shift $\sigma (\lambda_0)$ (Fig. \ref{SF_WithFeed_general}(b)) and the feedback strength $\eta_{F}$ (Fig. \ref{SF_WithFeed_general}(c)), related to the configurations with the lower error in Fig. \ref{SF_WithFeed_general}(a). 

\noindent While all the displayed configurations achieve errors lower than $NMSE_{SR}$, a joint evaluation of the three figures suggests the mechanisms behind the choice of the nonlinearity leading to these performance. At $P_{max}=0.1mW$, the MRR works in a linear regime, as also indicated by the small values of $\sigma (\lambda_0)$ at this power (Fig. \ref{SF_WithFeed_general}(b)). In this condition the feedback strength $\eta_{F}$ is maximized (Fig. \ref{SF_WithFeed_general}(c), at $P_{max}=0.1mW$). Doing so, the recursivity of the feedback signal in the system is increased and, consequently, the detection nonlinearity acts on a larger number of feedback delayed terms. This suggests that when the MRR is forced in a linear regime, due to the limited input optical power, the system enhances the linear memory using higher feedback strengths, to effectively solve the task. The best performance for this processing scheme results $NMSE=0.042 \pm 0.008$. At higher maximum incident power $P_{max}>0.1mW$, the MRR nonlinearity becomes also accessible, but still not mandatory. As illustrated in the previous task, the feedback phase $\Delta\phi_{F}$ can still be tuned such to minimize the optical power inside the MRR (destructive interference), and thus minimize its nonlinearity. Nevertheless, Fig. \ref{SF_WithFeed_general}(b) shows that for these incident optical powers, the standard deviation of the resonance shift, $\sigma (\lambda_0)$, is higher with respect to the linear case (Fig. \ref{SF_WithFeed_general}(b), at $P_{max}=0.1mW$), so that the system is actually exploiting the MRR nonlinearity. In particular, a region of lower NMSEs in Fig. \ref{SF_WithFeed_general}(a) is related to intermediate values of $\sigma (\lambda_0)$. In parallel, the strength of the feedback is reduced with respect to the linear case (Fig. \ref{SF_WithFeed_general}(c)). These results suggest that, under these conditions, the MRR contributes to the overall nonlinearity of the system and improves the prediction performance. Alternatively, the detection nonlinearity is also sufficient to solve the task, once enhanced by a stronger external optical feedback. 

\begin{figure}[htbp]
\centering\includegraphics[scale=0.35]{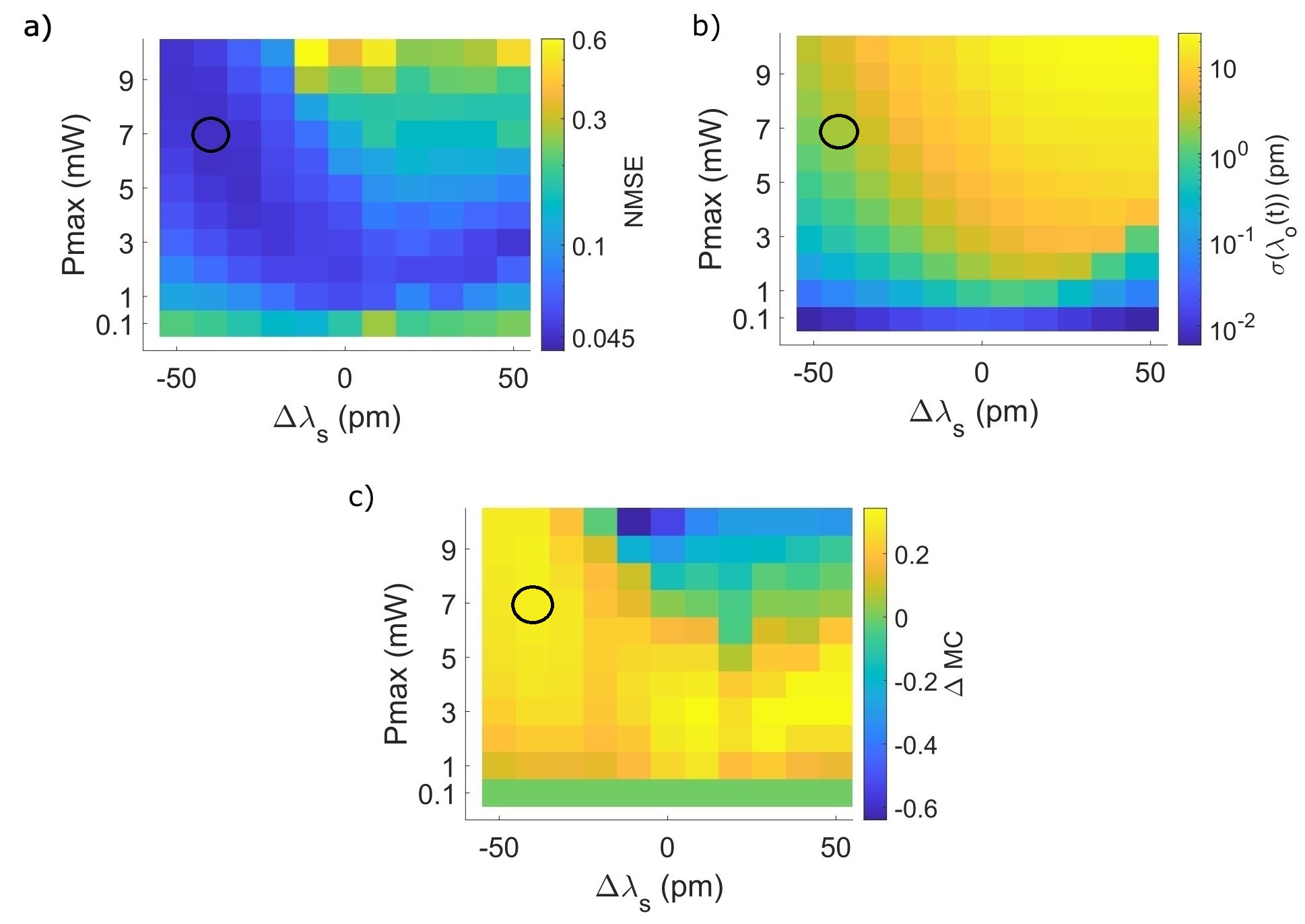}
\caption{Performance of the Santa Fe benchmark task, by using a single MRR without external feedback. (a) NMSE, (b)  standard deviation of the resonance wavelength shift $\sigma (\lambda_0)$, and (c) change of MC ($\Delta MC$), versus the starting wavelength shift $\Delta \lambda_s$ and the optical power $P_{max}$. The black circle indicates the conditions for which we obtain the lowest NMSE.}
\label{SF_NoFeed}
\end{figure}

The previously identified parameter set is not the only one that results in low NMSEs. An interesting operating condition is when we eliminate the external feedback cavity and evaluate the performance of the single MRR, by only adjusting its operating parameters $\Delta\lambda_{s}$ and $P_{max}$. In Fig. \ref{SF_NoFeed}(a) we show the NMSE performance, versus these two MRR parameters. The lowest error we obtain is $NMSE_{min} = 0.045\pm 0.002$, and it is achieved for $P_{max} = 7$ mW and $\Delta\lambda_{s} = -40$ pm (Fig. \ref{SF_NoFeed}(a), black circle). This NMSE value is only slightly higher than the one obtained from the MRR system with external optical feedback, indicating that the contribution of the extended external memory is not so critical for this task. To investigate further the conditions that lead to this performance, we map the standard deviation of the resonance wavelength shift $\sigma (\lambda_0)$, which indicates the strength of the MRR nonlinearity (Fig. \ref{SF_NoFeed}(b)), and the change in the $MC$ ($\Delta MC$), as shown in Fig. \ref{SF_NoFeed}(c). Here $\Delta MC = MC - MC_{lin}$, where $MC_{lin}$ is the linear memory capacity when the MRR operates in the linear regime. In both plots, the investigated parameter space is again $P_{max}$ versus $\Delta\lambda_{s}$. When the MRR works in the linear regime (i.e. for low optical power $P_{max} = 0.1$ mW), the NMSE we obtain is as high as 0.13 (Fig. \ref{SF_NoFeed}(a)). In this case, the MRR has only access to an inertia memory of one step, while the output undergoes a nonlinear transformation through the square-law photodetection. This is the reason why we obtain a $NMSE$ value lower than the $NMSE_{SR}$. By increasing $P_{max}$ and the nonlinear contribution of the MRR, the $NMSE$ is significantly reduced (Fig. \ref{SF_NoFeed}(a)). However, this happens only for those $\Delta\lambda_{s}$ conditions that preserve the free carrier nonlinearity of the MRR in a specific range of values (Fig. \ref{SF_NoFeed}(b)). Specifically, the $\sigma(\lambda_{o}$) should be not very low - which would mean an absence of nonlinear effects - but also less than few pm. In addition, we observe that these conditions also favor a higher $MC$ (Fig. \ref{SF_NoFeed}(c)). For conditions that lead to self pulsation dynamics (e.g. for high $P_{max}$, detuning dependent), the capability of the system to retain the memory is lower, the nonlinearity of the system is very high, and the overall capability for computation becomes limited. These findings are in agreement with \cite{Borghi_MRR}, where a single MRR in the absence of feedback is able to solve both the 1-bit delayed XOR task and the classification of Iris flowers, exploiting a combination of free-carrier and thermal nonlinear memories.

\section{Discussion}
In the RC implementation we propose, we process one piece of input information per time delay ($\tau_{F}$), while exploiting the free-carrier nonlinearity of the MRR. By considering in this study only 25 virtual nodes and a temporal spacing of $\theta = 40$ ps, the time delay of the feedback cavity is $\tau_{F}= 1$ ns. In tasks that might require a significantly higher number of nodes, this can be achieved by increasing the feedback delay up to several ns. In parallel, the temporal spacing $\theta$ can be further reduced. For example, with $\tau_{F} = 5$ ns and $\theta = 20$ ps, one can define 250 virtual nodes in the time-delay RC scheme, while still exploiting the free-carrier nonlinearity of the MRR (here $\tau_{FC} \approx 3$ ns). But, as the delay time becomes larger, the thermal effects contribution to the MRR nonlinearity also increases, as more free carriers are generated per input pulse. Thermal effects may either degrade the computing performance, or contribute positively by acting as a secondary source of large-scale nonlinear memory. In another consideration, one can exploit the MRR nonlinearity through only the thermal effects. The time delay can be then further expanded to  $\tau_{F} = 100$ ns. However, the task performances that we presented in this study will differ, due to the different nonlinear signal transformations. By changing the design and material parameters of the MRR, we are able to change the dynamical properties of the system. For example, an increase of the free carrier lifetime by silicon doping \cite{Fossum1983}, allows a larger $\tau_{F}$. In a RC topology, this translates into a larger number of virtual nodes that exploit the free carrier nonlinearity. For example, in  \cite{Borghi_MRR}, an unusually large recombination free carrier lifetime ($\tau_{FC} = 45$ ns) was exploited for computation. On the other hand, a reduction of the free carrier lifetime increases the processing speed, and can be realized with a p-i-n junction embedded in the MRR \cite{Xu2005}. The limited number of virtual nodes defined in this case can be partially compensated by adopting a smaller $\theta$. \\From an experimental point of view, for short delays – e.g. the one discussed above with $\tau_{F} = 5$ ns – an integrated solution with the MRR can be considered. For longer delays – e.g. the one discussed above with $\tau_{F} = 100$ ns – which are beyond the photonic integration capabilities, an optical fiber that is coupled to the MRR’s “through” and “add” ports may be considered. In both cases, for getting access to strong feedback conditions, an optical amplification unit is required in the feedback path to compensate for the coupling and propagation losses. The introduced optical noise in this case might have an effect on the final performance. An optical attenuator may be also used to further tune the feedback strength at lower values. Finally, a critical parameter to control is the feedback signal's phase ($\Delta\phi_{F}$). This can be experimentally realized via fine tuning of the laser emission wavelength or by using a piezo-driven phase-shifter within the feedback path. In fiber-based feedback delay lines, it is necessary to compensate thermal variations and mechanical vibrations, which may otherwise lead to phase drift in time. A PID controller can be used for this purpose \cite{Vinckier2015}.

\section{Conclusions}
Here we investigated numerically the capability of a passive silicon MRR with delayed optical feedback to operate as a versatile computational unit in time-delay RC. This work is in line with other time-delayed passive reservoirs already studied \cite{Vinckier2015,SESAM_2014}. While in those works the virtual nodes were coupled exploiting a temporal mismatch between the mask duration and the cavity delay, in our approach the virtual nodes are coupled via both the MRR dynamics and the feedback connectivity. MRRs have very small footprint, which makes them ideal candidates for scalable photonic RC integrated configurations. Designs with multiple coupled elements appear very prominent for enhanced dynamical response \cite{Mancinelli2014}. Moreover, by exploiting different resonant conditions, wavelength division multiplexing can be also supported. In the present study, we exploited the free carrier nonlinearity timescales, by selecting an external cavity delay and an encoding information duration of the same time scale (1 ns). This delay can be tuned to lower values to speed up computations, or to higher values to increase the number of virtual nodes. To test the computational properties of MRR in time-delay RC, we computed various benchmark tasks that have memory requirements. We showed that Narma 10 task can be solved efficiently, by operating the MRR in a linear regime and in presence of strong feedback, while using photodetection as the only source of signal nonlinear transformation. For the Mackey-Glass prediction task, we exploited both the MRR nonlinearity and the external cavity memory, to obtain the lowest prediction error. Finally, for the Santa Fe prediction task, we demonstrated that the MRR nonlinearity acts as a sufficient source of memory, eliminating the need for the external cavity, in agreement with the experimental results found in \cite{Borghi_MRR}.

\section*{Appendix: Parameters for numerical modeling the MRR system}
Most of the ring parameters we consider are reported in \cite{Matt_PhD}. $p$ is the MRR perimeter, $k^2$ is the power coupling coefficient between the MRR and the straight waveguides (here set to $0.01$, which is slightly lower than \cite{Matt_PhD}, to improve the enhancement factor and increase the nonlinear effects of the MRR), $a_{rt}$ is the MRR roundtrip field transmission coefficient, $Q$ is the MRR quality factor, $dn_{Si}/dT$ is the thermo-optic coefficient in silicon at $300 K$, $dn_{Si}/dN$ is the free carrier dispersion coefficient, $V_{eff}$ is the mode effective volume, $\beta_{TPA}$ is the TPA coefficient of silicon, and $P_{abs}$ is the total absorbed power (which includes linear absorption, TPA and FCA) within the MRR. The corresponding parameters used in the numerical simulations are provided in Table \ref{tab:parameters}. 

\begin{table}[]
    \centering
        \caption{Parameter values used in the numerical simulations in the model of Section \ref{sec_Model}.}
    \begin{tabular}{c|c|c|c}
      Parameter   & Value   &   parameter   &   Value\\
      \hline
       $p$    &    $2 \pi \times 6.75 \mu m$ &    $\lambda_{o}$    &    $1549.66 nm$ \\
       $\gamma_{i}$  &  $1.68 GHz$  &   $\gamma_{e}$   &   $17.2 GHz$ \\
       $k^2 = \frac{2 \Gamma p n_g}{c}$     &   0.01 ($n_g$=4.1) &  $t_{r}^2=1-k^2$    & 0.99 \\ 
       $a_{rt}$  & $e^{-\frac{c p}{4 n_g \gamma_{i}}}$    &   Q  & $\frac{\pi n_g p t_{r} \sqrt{a_{rt}} }{(1-t_{r}^2 a_{rt})\lambda_{o}}$ \\
       $dn_{si}/dT$    &    $1.86\times 10^{-4} K^{-1}$ &   $dn_{si}/dN$    &    $-4.2\times 10^{-27} m^3$ \\
       $n_{F}$    &    $1.4682$ &   $\tau_{ph}$   &    $52.81 ps$ \\  
       $\tau_{TH}$    &     $83.3 ns$   &   $\tau_{FC}$    &    $3.3 ns$    \\   
       $\Gamma_c$   &    $0.9$  &   $V_{eff}$    &   $5.331\times 10^{-18} m^3$\\   
       $\sigma_{FCA}$   &  $1.45\times 10^{-21} m^2$    &   $\eta_{FCA}$ &   $\frac{\sigma_{FCA} \Gamma_c c}{2 n_{Si}}$\\     
       $\beta_{TPA}$    &     $0.79\times 10^{-11} m/W$  &
       $P_{abs}$    &   $2\gamma(t)\left|U(t)\right|^2$ \\   
       $G_{TPA}$    &   $\frac{c^2 \beta_{TPA}}{2 V_{eff} n_{Si}^2}$      & $FWHM$    &  $\frac{\lambda_{o}}{Q}$\\
       
\end{tabular}
    \label{tab:parameters}
\end{table}

\vspace{1em}

\section*{Funding}

The work of G.D., C.R.M., and A.A. was partially supported by the Severo Ochoa and Maria de Maeztu Program for Centers and Units of Excellence in R$\&$D (MDM-2017-0711). The work of C.R.M and A.A. was also partially supported by the Ministerio de Ciencia e Innovación, and the projects PID2019-111537GB-C21 and PID2019-111537GB-C22. G.D., M.M., and L.P. also acknowledge the funding support of this work under the European Research Council (ERC) and the European Union’s Horizon 2020 research and innovation programme (Grant Agreement No. 788793, BACKUP).

\section*{Disclosures}

The authors declare that there are no conflicts of interest related to this article.

\section*{Data availability}

Data underlying the results presented in this paper are not publicly available at
this time but may be obtained from the authors upon reasonable request.

\bibliography{references}

\begin{thebibliography}{10}
\newcommand{\enquote}[1]{``#1''}

\bibitem{Heebner}
J.~Heebner, R.~Grover, and T.~Ibrahim, \emph{Optical microresonators: Theory,
  Fabrication, and Applications} (Springer Series in Optical Sciences).

\bibitem{Little1997}
B.~E. Little, S.~T. Chu, H.~A. Haus, Foresi, and J.~P. Laine,
  \enquote{Microring resonator channel dropping filters,}
  {\protect\JournalTitle{J. Lightw. Technol.}} \textbf{15}, 998–1005 (1997).

\bibitem{Cheng2018}
Q.~Cheng, S.~Rumley, M.~Bahadori, and K.~Bergman, \enquote{Photonic switching
  in high performance datacenters,} {\protect\JournalTitle{Opt. Express}}
  \textbf{26}, 16022--16043 (2018).

\bibitem{mesaritakis2010}
C.~Mesaritakis, A.~Argyris, E.~Grivas, A.~Kapsalis, and D.~Syvridis,
  \enquote{Adaptive interrogation for fast optical sensing based on cascaded
  micro-ring resonators,} {\protect\JournalTitle{IEEE Sensors Journal}}
  \textbf{11}, 1595--1601 (2010).

\bibitem{Sensing2019}
P.~Steglich, M.~Hülsemann, B.~Dietzel, and A.~Mai, \enquote{Optical biosensors
  based on silicon-on-insulator ring resonators: A review,}
  {\protect\JournalTitle{Molecules}} \textbf{24}, 519 (2019).

\bibitem{Ferrara2010}
M.~Ferrara, Y.~Park, L.~Razzari, B.~E. Little, S.~T. Chu, R.~Morandotti, D.~J.
  Moss, and J.~Azaña, \enquote{On-chip cmos-compatible all-optical
  integrator,} {\protect\JournalTitle{Nature Communications}} \textbf{1}, 29
  (2010).

\bibitem{Xu2005}
Q.~Xu, B.~Schmidt, S.~Pradhan, and M.~Lipson, \enquote{Micrometre-scale silicon
  electro-optic modulator,} {\protect\JournalTitle{Nature}} \textbf{435},
  325–327 (2005).

\bibitem{Zhang2013}
L.~Zhang, Y.~Fei, T.~Cao, Y.~Cao, Q.~Xu, and S.~Chen, \enquote{Multibistability
  and self-pulsation in nonlinear high-q silicon microring resonators
  considering thermo-optical effect,} {\protect\JournalTitle{Phys. Rev. A}}
  \textbf{87}, 53805 (2013).

\bibitem{Johnson2006}
T.~J. Johnson, M.~Borselli, and O.~Painter, \enquote{Self-induced optical
  modulation of the transmission through a high-q silicon microdisk resonator,}
  {\protect\JournalTitle{Opt. Express}} \textbf{14}, 817--831 (2006).

\bibitem{Almeida2004}
V.~R. Almeida and M.~Lipson, \enquote{Optical bistability on a silicon chip,}
  {\protect\JournalTitle{Opt. Lett.}} \textbf{29}, 2387–2389 (2004).

\bibitem{Preble2005}
S.~F. Preble, Q.~Xu, B.~S. Schmidt, and M.~Lipson, \enquote{Ultrafast
  all-optical modulation on a silicon chip,} {\protect\JournalTitle{Opt.
  Lett.}} \textbf{30}, 2891--2893 (2005).

\bibitem{Xu2007}
Q.~Xu and M.~Lipson, \enquote{All-optical logic based on silicon micro-ring
  resonators,} {\protect\JournalTitle{Opt. Express}} \textbf{15}, 924--929
  (2007).

\bibitem{Vaerenbergh2012}
T.~Van~Vaerenbergh, M.~Fiers, P.~Mechet, T.~Spuesens, R.~Kumar, G.~Morthier,
  B.~Schrauwen, J.~Dambre, and P.~Bienstman, \enquote{Cascadable excitability
  in microrings,} {\protect\JournalTitle{Opt. Express}} \textbf{20},
  20292--20308 (2012).

\bibitem{Xiang2020}
J.~Xiang, A.~Torchy, X.~Guo, and Y.~Su, \enquote{All-optical spiking neuron
  based on passive microresonator,} {\protect\JournalTitle{J. Lightw.
  Technol.}} \textbf{38}, 4019--4029 (2020).

\bibitem{ESN_2004}
H.~Jaeger and H.~Haas, \enquote{Harnessing nonlinearity: Predicting chaotic
  systems and saving energy in wireless communication,}
  {\protect\JournalTitle{Science}} \textbf{304}, 78--80 (2004).

\bibitem{LSM_2002}
W.~Maas, T.~Natschläger, and H.~Markram, \enquote{Real-time computing without
  stable states: a new framework for neural computation based on
  perturbations,} {\protect\JournalTitle{Neural Comput.}} \textbf{14},
  2531–2560 (2002).

\bibitem{MR_RC}
M.~Denis-Le~Coarer, Sciamanna, A.~Katumba, M.~Freiberger, J.~Dambre,
  P.~Bienstman, and D.~Rontani, \enquote{All-optical reservoir computing on a
  photonic chip using silicon-based ring resonators,}
  {\protect\JournalTitle{IEEE Journal of Selected Topics in Quantum
  Electronics}} \textbf{24}, 1--8 (2018).

\bibitem{Li2021}
S.~Li, S.~Dev, S.~Kühl, K.~Jamshidi, and S.~Pachnicke, \enquote{Micro-ring
  resonator based photonic reservoir computing for pam equalization,}
  {\protect\JournalTitle{IEEE Photonics Technology Letters}} \textbf{33},
  978--981 (2021).

\bibitem{Appeltant_2011}
L.~Appeltant, M.~C. Soriano, G.~Van~der Sande, J.~Danckaert, S.~Massar,
  J.~Dambre, B.~Schrauwen, C.~R. Mirasso, and I.~Fischer, \enquote{Information
  processing using a single dynamical node as complex system,}
  {\protect\JournalTitle{Nature Communications}} \textbf{2}, 468 (2011).

\bibitem{Larger2012}
L.~Larger, M.~C. Soriano, D.~Brunner, L.~Appeltant, J.~M. Gutierrez,
  L.~Pesquera, C.~R. Mirasso, and I.~Fischer, \enquote{Photonic information
  processing beyond turing: an optoelectronic implementation of reservoir
  computing,} {\protect\JournalTitle{Opt. Express}} \textbf{20}, 3241–3249
  (2012).

\bibitem{Brunner_Laser}
D.~Brunner, M.~C. Soriano, C.~R. Mirasso, and I.~Fischer, \enquote{Parallel
  photonic information processing at gigabyte per second data rates using
  transient states,} {\protect\JournalTitle{Nat Commun}} \textbf{4}, 1364
  (2013).

\bibitem{laserNumeric2014}
R.~M. Nguimdo, G.~Verschaffelt, J.~Danckaert, and G.~Van~der Sande,
  \enquote{Fast photonic information processing using semiconductor lasers with
  delayed optical feedback: Role of phase dynamics,}
  {\protect\JournalTitle{Opt. Express}} \textbf{22}, 8672--8686 (2014).

\bibitem{Bueno2017}
J.~Bueno, D.~Brunner, M.~C. Soriano, and I.~Fischer, \enquote{Conditions for
  reservoir computing performance using semiconductor lasers with delayed
  optical feedback,} {\protect\JournalTitle{Opt. Express}} \textbf{25},
  2401–2412 (2017).

\bibitem{Takano2018}
K.~Takano, C.~Sugano, M.~Inubushi, K.~Yoshimura, S.~Sunada, K.~Kanno, and
  A.~Uchida, \enquote{Compact reservoir computing with a photonic integrated
  circuit,} {\protect\JournalTitle{Opt. Express}} \textbf{26}, 29424–29439
  (2018).

\bibitem{Harkhoe2020}
K.~Harkhoe, G.~Verschaffelt, A.~Katumba, P.~Bienstman, and G.~Van~der Sande,
  \enquote{Demonstrating delay-based reservoir computing using a compact
  photonic integrated chip,} {\protect\JournalTitle{Opt. Express}} \textbf{28},
  3086--3096 (2020).

\bibitem{Borghi_MRR}
M.~Borghi, S.~Biasi, and L.~Pavesi, \enquote{Reservoir computing based on a
  silicon microring and time multiplexing for binary and analog operations,}
  {\protect\JournalTitle{Sci Rep}} \textbf{11}, 15642 (2021).

\bibitem{NLvsLMemory2017}
M.~Inubushi and K.~Yoshimura, \enquote{Reservoir computing beyond
  memory-nonlinearity trade-off,} {\protect\JournalTitle{Scientific rep.}}
  \textbf{7}, 10199 (2017).

\bibitem{Jaeger_MC}
H.~Jaeger, \enquote{Short term memory in echo state networks,}
  {\protect\JournalTitle{GMD Report}} \textbf{152}, 60 (2002).

\bibitem{Atiya2000}
A.~F. Atiya and A.~G. Parlos, \enquote{New results on recurrent network
  training: unifying the algorithms and accelerating convergence,}
  {\protect\JournalTitle{I EEE T. Neural Netw.}} \textbf{11}, 697–709 (2000).

\bibitem{Vinckier2015}
Q.~Vinckier, F.~Duport, A.~Smerieri, K.~Vandoorne, P.~Bienstman, M.~Haelterman,
  and S.~Massar, \enquote{High-performance photonic reservoir computer based on
  a coherently driven passive cavity,} {\protect\JournalTitle{Optica}}
  \textbf{2}, 438--446 (2015).

\bibitem{MG1977}
M.~C. Mackey and L.~Glass, \enquote{Oscillation and chaos in physiological
  control systems,} {\protect\JournalTitle{Science}} \textbf{197}, 287--289
  (1977).

\bibitem{SF93}
A.~S. Weigend and N.~A. Gershenfeld, \enquote{Results of the time series
  prediction competition at the santa fe institute,}
  {\protect\JournalTitle{IEEE International Conference on Neural Networks}}
  \textbf{3}, 1786–1793 (1993).

\bibitem{Fossum1983}
J.~G. Fossum, R.~P. Mertens, D.~S. Lee, and J.~F. Nijs, \enquote{Carrier
  recombination and lifetime in highly doped silicon,}
  {\protect\JournalTitle{Solid-State Electronics}} \textbf{26}, 569--576
  (1983).

\bibitem{SESAM_2014}
A.~Dejonckheere, F.~Duport, A.~Smerieri, L.~Fang, J.~L. Oudar, M.~Haelterman,
  and S.~Massar, \enquote{All-optical reservoir computer based on saturation of
  absorption,} {\protect\JournalTitle{Opt. Express}} \textbf{22}, 10868--10881
  (2014).

\bibitem{Mancinelli2014}
M.~Mancinelli, M.~Borghi, F.~Ramiro-Manzano, J.~M. Fedeli, and L.~Pavesi,
  \enquote{Chaotic dynamics in coupled resonator sequences,}
  {\protect\JournalTitle{Opt. Express}} \textbf{22}, 14505--14516 (2014).

\bibitem{Matt_PhD}
M.~Mancinelli, \enquote{Linear and non linear coupling effects in sequence of
  microresonators,} Ph.D. thesis, University of Trento (2013).

\end{thebibliography}

\end{document}